\documentclass[final]{JHEP3} 

\JHEPspecialurl{http://jhep.sissa.it/JOURNAL/JHEP3.tar.gz}

\usepackage{epsfig,multicol,bbm,amsmath,amssymb}

\newcommand\fverb{\setbox\fverbbox=\hbox\bgroup\verb}
\newcommand\fverbdo{\egroup\medskip\noindent%
			\fbox{\unhbox\fverbbox}\ }
\newcommand\fverbit{\egroup\item[\fbox{\unhbox\fverbbox}]}
\newbox\fverbbox

\def\ra{\rightarrow}

\def\chioi{\tilde{\chi}^0_1}
\def\chioii{\tilde{\chi}^0_2}

\def\etm{E_T^{\rm miss}}
\def\ra{\rightarrow}

\def\slep{\tilde{\ell}}

\def\mt{M_T}
\def\mtmax{M_T^{\rm max}}
\def\mmax{m_{\rm max}}
\def\mmin{m_{\rm min}}

\title{Transverse mass and invariant mass observables for measuring
the mass of a semi-invisibly decaying heavy particle.}

\author{Daniel R. Tovey,\\ Department of Physics and
	Astronomy, \\ University of Sheffield, Hounsfield Road,
	Sheffield S3 7RH, UK\\ E-mail:\email{ daniel.tovey@cern.ch}}

\received{} 		
\accepted{}		

\preprint{}

\abstract{Formulae are derived for the positions of end-points in the
invariant mass and transverse mass distributions obtained from the
products of heavy states decaying to pairs of semi-invisibly decaying
lighter states. Formulae are derived both for the special case where
the two decay chains are identical and the more general case where
they are different. The formulae are tested with a simple case study
of heavy SUSY higgs particles decaying to gauginos at the LHC.}


\begin{document} 


\section{Introduction}\label{sec1}

In recent years there has been considerable theoretical and
experimental interest in the possibility of observing and measuring
the production of heavy neutral SUSY higgs particles ($H/A$) at the
LHC. In many regions of parameter space decays to Standard Model (SM)
particles such as $H/A \ra \tau\tau$ provide the clean signature of a
peak at $m(H/A)$ in the reconstructed invariant mass distribution of
the decay products. In certain regions of parameter space however,
particularly at larger values of $m_A$, this channel is suppressed and
alternative signatures must be sought. One interesting possibility
\cite{Baer:1992kd,Baer:1994fx,Bisset:1995dc,Moortgat:2001pp,Bisset:2003ix,Nojiri:2003tu,Charlot:2006se,Bian:2006xx,Bisset:2007mi,Bisset:2007mk,Huang:2008qd,Gentile:2009zz,Gentile:2000zz}
involves the use of higgs decays to pairs of gauginos (neutralinos or
charginos), with subsequent gaugino decays to dileptons and the
Lightest Supersymmetric Particle (LSP -- here assumed to be neutral
and stable). The resulting $4\ell+\etm$ signature suffers from
relatively little irreducible Standard Model background contamination,
although there can be significant irreducible SUSY background from
direct or indirect gaugino production (the latter via squark decays).

Several detailed studies have been performed of the likely sensitivity
of the LHC detectors to SUSY higgs production using the $4\ell+\etm$
signature (see
e.g. Refs.~\cite{Baer:1992kd,Baer:1994fx,Bisset:1995dc,Moortgat:2001pp,Bisset:2003ix,Nojiri:2003tu,Charlot:2006se,Bian:2006xx,Bisset:2007mi,Bisset:2007mk,Huang:2008qd,Gentile:2009zz,Gentile:2000zz}). If
a statistically significant excess is observed in this channel then it
is natural to ask whether the same channel can be used to measure the
mass of the higgs. Unlike in the case of SM decays it is likely to be
extremely difficult to reconstruct a higgs mass peak due to the
presence of the two heavy invisible LSPs in each higgs decay. In the
case where the decay proceeds through the chain:
\begin{equation}
\label{eqn1}
H/A \ra \chioii\chioii \ra \ell\slep \ell\slep \ra \ell\ell\ell\ell \chioi\chioi,
\end{equation}
and the masses of the $\chioii$, $\slep$ and $\chioi$ (LSP) are known,
it is possible to solve the kinematic constraints to obtain an
event-by-event value of $m(H/A)$ \cite{Nojiri:2003tu}. If the sleptons
$\slep$ are heavy however, causing the $\chioii$ particles decay
through the three-body process $\chioii \ra \ell \ell \chioi$, there
are insufficient constraints to solve for $m(H/A)$. In this case one
must resort to measuring the positions of end-points in the
distributions of event invariant mass and transverse mass
\cite{Arnison:1983rp,Banner:1983jy,Arnison:1983zy,Gripaios:2007is,Barr:2007hy,Barr:2009mx,Barr:2009jv}
values, which depend on the masses of all the particles involved in
the decay, including $m(H/A)$.

In this brief paper we shall derive formulae for the positions of these
end-points both in the special case where the decay chains are
identical and in the more general case of non-identical decay
chains. The resulting expressions will be applicable to any channel in
which a heavy state $\omega$ decays to a pair of lighter states
$\delta_i$ ($i=1,2$), each of which decays in turn to an aggregate
visible state $v_i$ and an invisible particle $\alpha_i$:
\begin{equation}
\label{eqn2}
\omega \ra \delta_1\delta_2 \ra v_1\alpha_1 v_2\alpha_2.
\end{equation}
Note that the aggregate visible states $v_i$ can each be composed of
multiple visible particles -- the $v_i$ should be considered to be
pseudo-particles with four-momenta equal to the net four-momenta of
their constituents. We shall test the new formulae with a simple
case-study of heavy SUSY higgs decays to gauginos. We shall not
attempt to perform a full detector-level study of the SM and SUSY
backgrounds to the $4\ell+\etm$ channel as this is described elsewhere
(see
e.g. Refs.~\cite{Moortgat:2001pp,Bisset:2007mi,Gentile:2009zz,Gentile:2000zz}),
however we shall discuss the relative merits of using the invariant
mass and transverse mass end-points to measure $m(H/A)$.

The structure of the paper is as follows. In Section~\ref{sec2} we
derive the invariant mass end-point formulae while in
Section~\ref{sec3} we derive the equivalent formulae for transverse
mass end-points. In Section~\ref{sec4} we present the results of the
case-study of SUSY higgs decays to four leptons and $\etm$. In
Section~\ref{sec5} we conclude.

\section{Invariant mass end-points}\label{sec2}

As is well-known the invariant mass of the aggregate visible products
of each chain, $v_1$ and $v_2$, is given by:
\begin{eqnarray}
\label{eqn3}
m^2(v_1,v_2) & = & [E(v_1)+E(v_2)]^2 - [{\bf p}(v_1)+{\bf p}(v_2)]^2 \cr
           & = & m^2(v_1)+m^2(v_2)+2[E(v_1)E(v_2)-{\bf p}(v_1) \cdot {\bf p}(v_2)],
\end{eqnarray}
where bold quantities denote three-momenta. The maximum value taken by
$m(v_1,v_2)$ depends upon whether the $\alpha_i$ particles can be
brought to rest in the $\omega$ rest frame. Such a configuration can
be obtained in principle if the boosts of the $\alpha_i$ particles in
the respective $\delta_i$ rest frames can be arranged to exactly
cancel the boosts they obtain in the $\omega$ rest frame from the
boosts of the $\delta_i$ rest frames. If this condition can be
satisfied for some values of $m(v_i)$ then this configuration
generates the maximum possible value of $m(v_1,v_2)$:
\begin{equation}
\label{eqn4}
\mmax(v_1,v_2) = m(\omega)-m(\alpha_1)-m(\alpha_2).
\end{equation}
The necessary and sufficient conditions imposed upon the masses by
this condition are discussed below for the cases where the two
$\delta_i$ decay chains are identical or different.

\subsection{Identical chains}\label{subsec2.1}

When $\delta_1$ and $\delta_2$ are of equal mass, and similarly for
$\alpha_1$ and $\alpha_2$, the condition that both $\alpha_i$
particles be brought to rest in the $\omega$ rest frame is equivalent
to the requirement:
\begin{equation}
\label{eqn5}
\mmax^2(v) \geq m^2(\delta)-m(\alpha)[m(\omega)-m(\alpha)] \geq \mmin^2(v),
\end{equation}
where $\mmax(v)$ and $\mmin(v)$ are respectively the maximum and
minimum possible invariant masses of all the visible decay products of
each chain. If this condition can be satisfied then from
Eqn.~(\ref{eqn4})
\begin{equation}
\label{eqn5a}
\mmax(v_1,v_2)=m(\omega)-2m(\alpha). 
\end{equation}
In models satisfying Eqn.~(\ref{eqn5}) the kinematic configuration
saturating the bound Eqn.~(\ref{eqn5a}) is that in which both
$\alpha_i$ particles are emitted against the motion of their parent
$\delta_i$ particles with sufficient momentum to cancel the boost in
the $\omega$ rest frame provided by this motion. 

\FIGURE[t]{
\epsfig{file=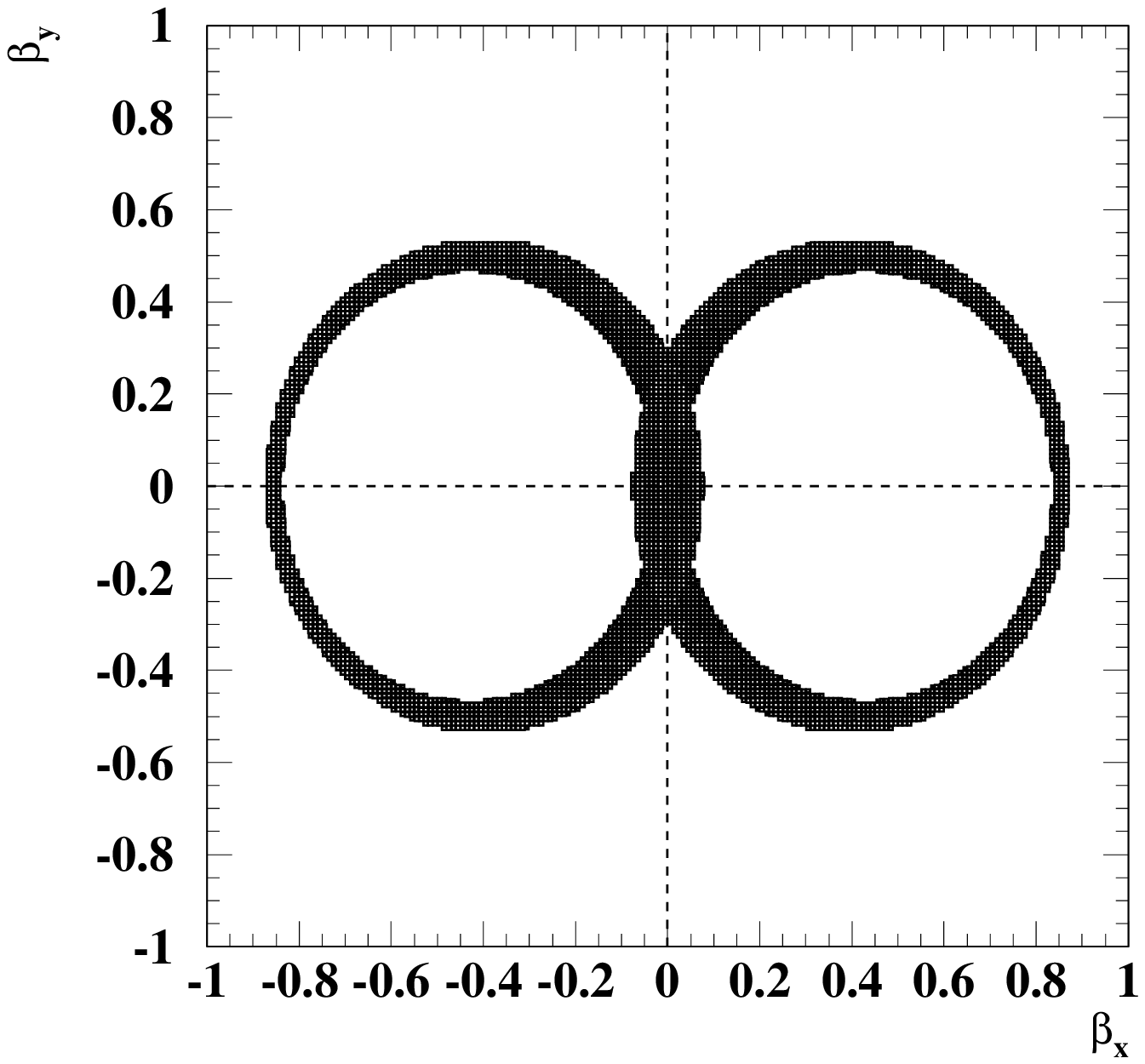,height=1.9in}
\epsfig{file=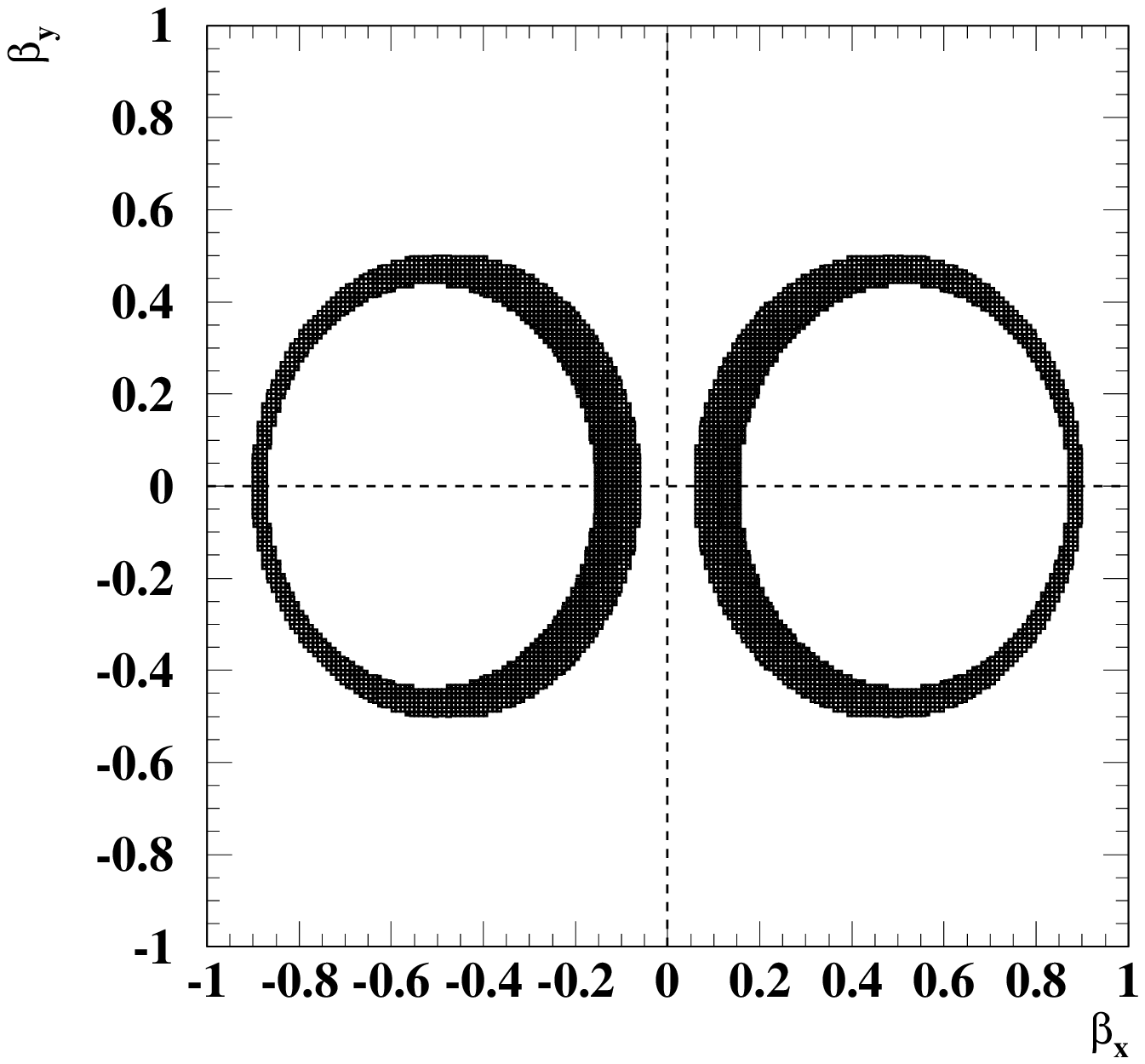,height=1.9in}
\epsfig{file=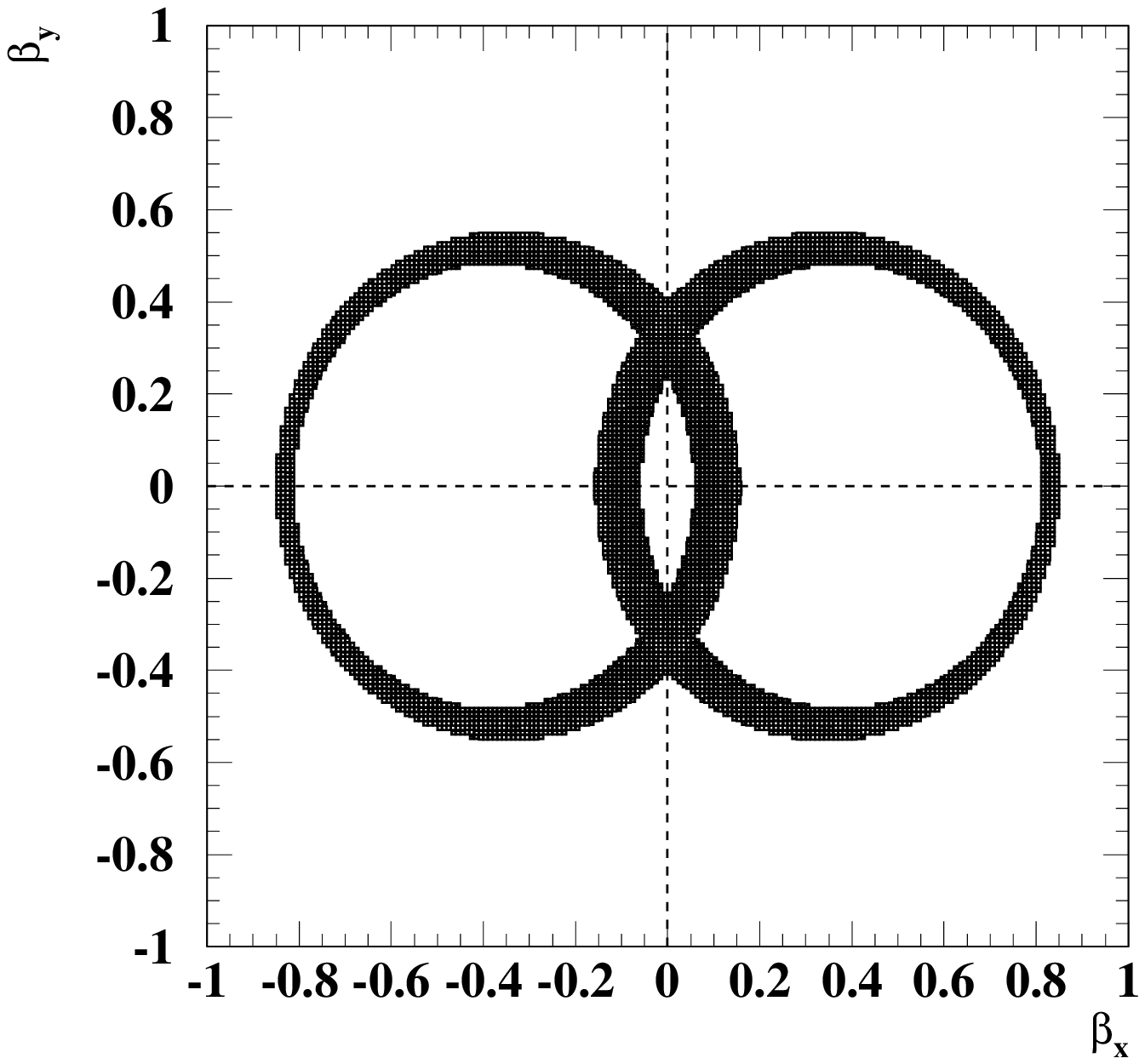,height=1.9in}
\caption{\label{fig1} $\alpha_i$ phase space diagrams for toy models
with identical decay chains. The hatched regions represent the annular
rings referred to in the text. See text for details of models.}}  A
useful means of visualising the kinematic configurations generated by
a given model is provided by phase-space diagrams in which the
cylindrical polar components of the velocities $\beta$ of the two
$\alpha_i$ particles measured in the $\omega$ rest frame are plotted
on the $\hat{x}$ axis (components parallel to the $\omega$ decay axis)
and $\hat{y}$ axis (components perpendicular to the $\omega$ decay
axis). Examples of such diagrams are shown in Figure~\ref{fig1}. In
these diagrams the phase space accessible to each $\alpha_i$ particle
is represented by an annular ring, with the outer and inner boundaries
defined by the kinematic configurations in which $m(v_i)=\mmin(v_i)$
and $m(v_i)=\mmax(v)$ respectively. If
$\mmax(v)=m(\delta_i)-m(\alpha_i)$, as is the case for three-body
$\delta_i$ decays, then the inner boundary possesses zero radius. This
is because in this case the $\alpha_i$ particles must be produced at
rest in the $\delta_i$ rest frames.

Figure~\ref{fig1}(left) shows the $\alpha_i$ phase space diagram for a
toy model satisfying Eqn.~(\ref{eqn5}). The configuration saturating
the bound Eqn.~(\ref{eqn5}) is that in which $\beta=0$ for both
$\alpha_i$ particles, which requires that the origin lies within both
annular regions.

Figure~\ref{fig1}(centre) and Figure~\ref{fig1}(right) show $\alpha_i$
phase space diagrams for models which do not satisfy
Eqn.~(\ref{eqn5}), as can easily be seen by observing that in neither
case does the origin lie within the annular regions. In such cases
$\mmax(v_1,v_2)$ is obtained from configurations in which
$m(\alpha_1,\alpha_2)$ is minimised. These configurations must
therefore simultaneously minimise the velocity $\beta$ of each
$\alpha_i$ particle in the $\omega$ rest frame. In models similar to
that shown in Figure~\ref{fig1}(centre) this requires that
$m(v_1)=m(v_2)=\mmin(v)$, generating configurations lying on the outer
boundaries of the annular regions. In Figure~\ref{fig1}(right) this
requires that $m(v_1)=m(v_2)=\mmax(v)$ generating configurations lying
on the inner boundaries. In both cases the configurations saturating
the bound generate $\beta$ values lying on the $\hat{x}$-axis. This
can be understood by remembering that the closest point to the origin
on a circle with centre displaced from the origin along a given axis
lies at the intersection between that axis and the circle.

In practice, if Eqn.~(\ref{eqn5}) is not satisfied for a given model
then $\mmax(v_1,v_2)$ can be determined from the maximum of the two
values of $m(v_1,v_2)$ obtained when $m(v_1)=m(v_2)=\mmax(v)$ and
$m(v_1)=m(v_2)=\mmin(v)$, assuming that both $\alpha_i$ particles
are emitted against the directions of motion of their parent
$\delta_i$ particles. For fixed $m(v_1)=m(v_2)=m(v)$ this
configuration leads to the following expression for $m(v_1,v_2)$:
\begin{multline}
\label{eqn6}
m(v_1,v_2) = \frac{m(\omega)}{2m^2(\delta)}\Big[m^2(\delta)-m^2(\alpha)+m^2(v)+\\\frac{1}{m(\omega)}\sqrt{\big(m^2(\omega)-4m^2(\delta)\big)\big(\big[m^2(\delta)-m^2(\alpha)+m^2(v)\big]^2-4m^2(\delta)m^2(v)\big)}\Big],
\end{multline}
and so substituting respectively $m(v)=\mmax(v)$ and
$m(v)=\mmin(v)$ into Eqn.~(\ref{eqn6}) enables $\mmax(v_1,v_2)$ to
be determined.

\subsection{Non-identical chains}\label{subsec2.2}

\FIGURE[t]{
\epsfig{file=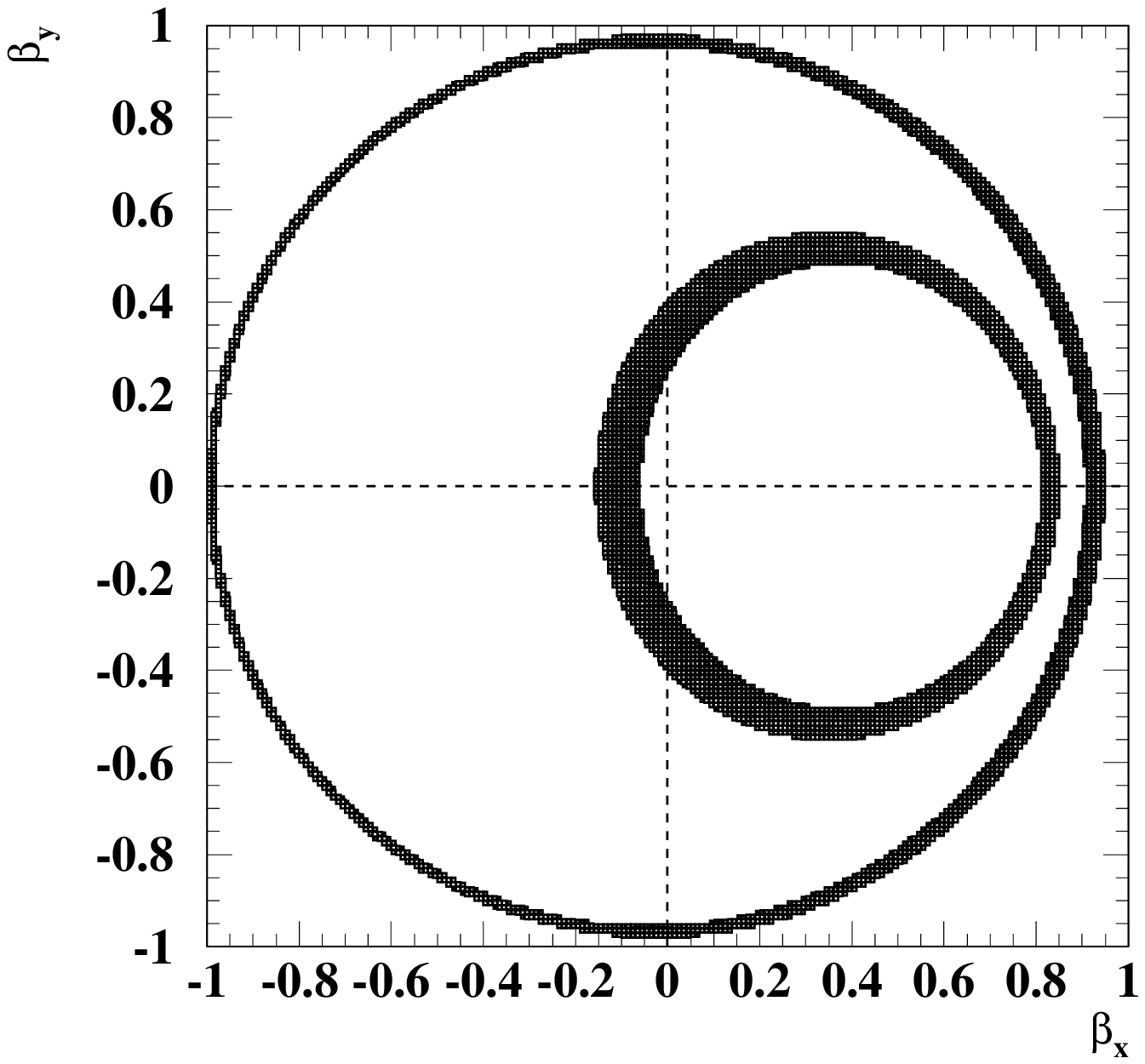,height=1.9in}
\epsfig{file=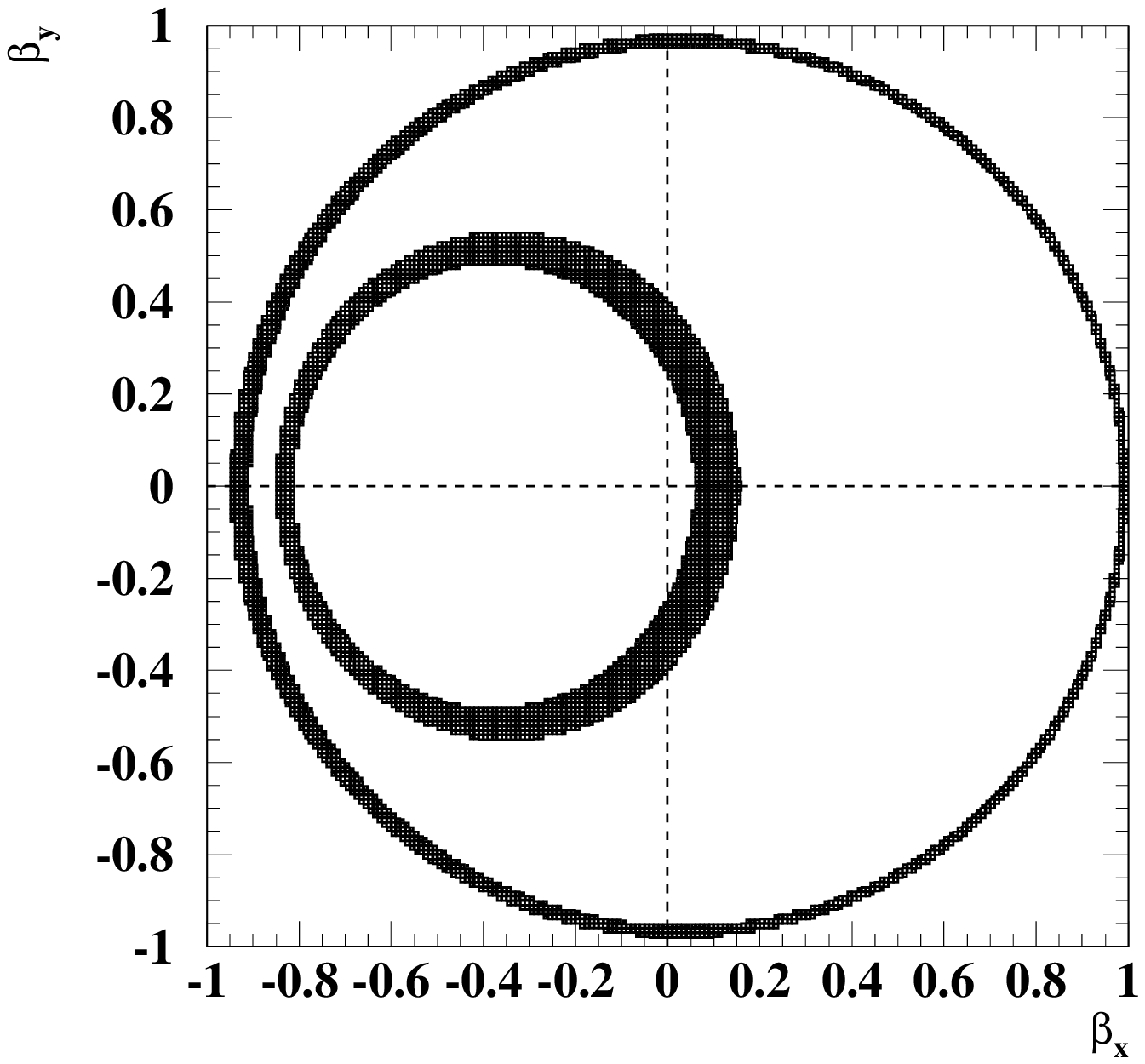,height=1.9in}
\epsfig{file=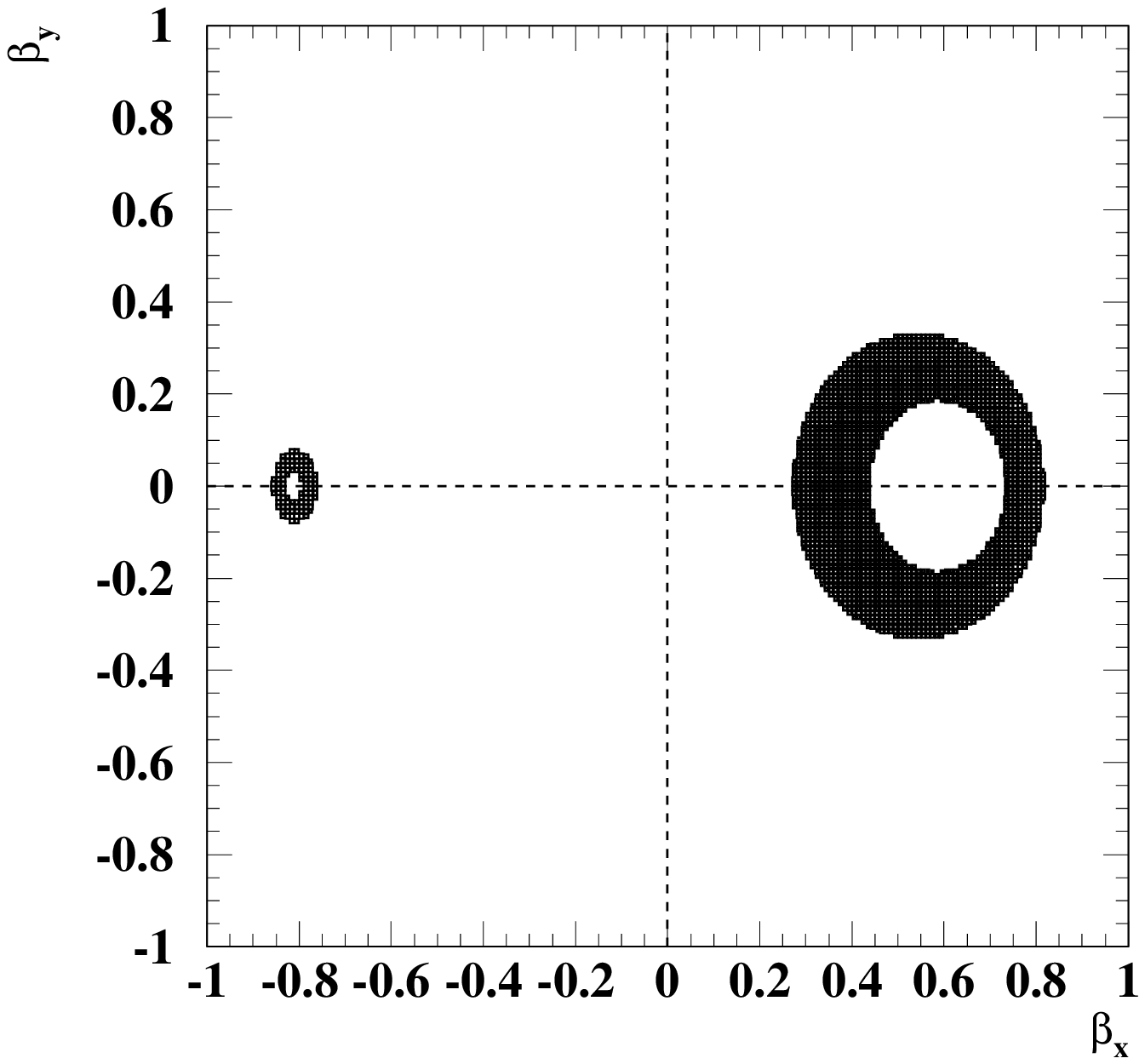,height=1.9in}
\caption{\label{fig2} $\alpha_i$ phase space diagrams for toy models with
non-identical decay chains. The left(right)-hand annular regions in
each figure correspond to decay chain 1(2). See text for details of
models.}}

When the two decay chains are not identical alternative kinematic
configurations can maximise $m(v_1,v_2)$. In such models the
$\alpha_i$ phase space diagrams are not in general symmetric and
qualitatively new configurations can be generated such as those
appearing in the diagrams in Figure~\ref{fig2}. In
Figure~\ref{fig2}(left) and Figure~\ref{fig2}(centre) the energy
released in the decay of $\delta_1$ is respectively less than or
greater than that released in the decay of $\delta_2$. In
Figure~\ref{fig2}(right) $m(\delta_2)$ is very much less than
$m(\delta_1)$ leading to a larger boost of $\delta_2$ and hence a
larger offset of the centre of the left-hand annular region. It should
be clear from this diagram that the number of kinematic configurations
which can in principle maximise $m(v_1,v_2)$ is greater than in the
case of identical decay chains.

As in the case of identical decay chains discussed above, the maximum
possible value of $m(v_1,v_2)$ is given by Eqn.~(\ref{eqn4}) if the
model can generate a configuration in which both $\alpha_i$ particles
are at rest in the $\omega$ rest frame. The conditions for this to be
the case for non-identical decay chains, analogous to Eqn.~(\ref{eqn5})
for identical decay chains are:
\begin{eqnarray}
\mmax^2(v_1) \geq & m^2(\delta_1)-m(\alpha_1)\Big[m(\omega)-m(\alpha_1)+\frac{m^2(\delta_1)-m^2(\delta_2)}{m(\omega)}\Big] & \geq \mmin^2(v_1), \label{eqn7-1} \\
\mmax^2(v_2) \geq & m^2(\delta_2)-m(\alpha_2)\Big[m(\omega)-m(\alpha_2)+\frac{m^2(\delta_2)-m^2(\delta_1)}{m(\omega)}\Big] & \geq \mmin^2(v_2).\label{eqn7-2}
\end{eqnarray}

To identify the possible configurations which can maximise
$m(v_1,v_2)$ when the conditions Eqns.~(\ref{eqn7-1})
and~(\ref{eqn7-2}) are not satisfied let us first consider the
energies and momenta of the $\alpha_i$ and $v_i$ particles in the
$\delta_i$ rest frames. These may be determined from simple two-body
kinematics to be:
\begin{eqnarray}
\label{eqn7}
E(\alpha_1) = \frac{m^2(\delta_1)+m^2(\alpha_1)-m^2(v_1)}{2m(\delta_1)},&\mbox{  }&
E(v_1) = \frac{m^2(\delta_1)-m^2(\alpha_1)+m^2(v_1)}{2m(\delta_1)}, \cr
E(\alpha_2) = \frac{m^2(\delta_2)+m^2(\alpha_2)-m^2(v_2)}{2m(\delta_2)},&\mbox{  }&
E(v_2) = \frac{m^2(\delta_2)-m^2(\alpha_2)+m^2(v_2)}{2m(\delta_2)}, 
\end{eqnarray}
and
\begin{eqnarray}
\label{eqn7a}
p(v_1) = p(\alpha_1) & = & \sqrt{E^2(v_1)-m^2(v_1)} = \sqrt{E^2(\alpha_1)-m^2(\alpha_1)}, \cr
p(v_2) = p(\alpha_2) & = & \sqrt{E^2(v_2)-m^2(v_2)} = \sqrt{E^2(\alpha_2)-m^2(\alpha_2)},
\end{eqnarray}
where $p(x)$ denotes the magnitude of the three-momentum of particle
$x$. In addition the energies and momenta of the $\delta_i$ particles
in the $\omega$ rest frame are given by:
\begin{eqnarray}
\label{eqn8}
E(\delta_1) = \frac{m^2(\omega)+m^2(\delta_1)-m^2(\delta_2)}{2m(\omega)},&\mbox{  }&
E(\delta_2) = \frac{m^2(\omega)-m^2(\delta_1)+m^2(\delta_2)}{2m(\omega)},
\end{eqnarray}
and
\begin{eqnarray}
\label{eqn8a}
p(\delta_1) = p(\delta_2) & = & \sqrt{E^2(\delta_1)-m^2(\delta_1)} = \sqrt{E^2(\delta_2)-m^2(\delta_2)}.
\end{eqnarray}
The magnitudes of the boosts $\beta_i$ provided to the $\delta_i$ particles by
the $\omega$ decay can be obtained from 
\begin{equation}
\label{eqn9}
\beta_i = \frac{p(\delta_i)}{E(\delta_i)}.
\end{equation}

Now let us consider the possible extremal values of the momenta and
energies of the $\alpha_i$ and $v_i$ particles in the $\omega$ rest
frame, for fixed $m(v_1)$ and $m(v_2)$. These can be obtained through
Lorentz transformations of Eqns.~(\ref{eqn7}) and~(\ref{eqn7a}) with
$\beta_i$ values obtained from Eqns.~(\ref{eqn8})--(\ref{eqn9}):
\begin{eqnarray}
\label{eqn10}
E'(\alpha_1) = \gamma_1[E(\alpha_1) + c_1 \beta_1 p(\alpha_1)], & \mbox{   } & p'(\alpha_1) = \gamma_1[c_1 p(\alpha_1) + \beta_1 E(\alpha_1)], \cr
E'(\alpha_2) = \gamma_2[E(\alpha_2) - c_2 \beta_2 p(\alpha_2)], & \mbox{   } & p'(\alpha_2) = \gamma_2[c_2 p(\alpha_2) - \beta_2 E(\alpha_2)], \cr
E'(v_1) = \gamma_1[E(v_1) - c_1 \beta_1 p(v_1)], & \mbox{   } & p'(v_1) = \gamma_1[ -c_1 p(v_1) + \beta_1 E(v_1)], \cr
E'(v_2) = \gamma_2[E(v_2) + c_2 \beta_2 p(v_2)], & \mbox{   } & p'(v_2) = \gamma_2[ -c_2 p(v_2) - \beta_2 E(v_2)],
\end{eqnarray}
where primed quantities are measured in the $\omega$ rest frame and
$\gamma_i$ is the Lorentz gamma factor derived from $\beta_i$. The
quantities $c_i=\pm 1$ and are determined by the directions in which
the $\alpha_i$ particles are emitted in the $\delta_i$ rest frames. If
we define $\delta_1$ to be travelling in the $+\hat{x}$ direction in
the $\omega$ rest frame then $c_1=+1$ corresponds to a configuration
in which $\alpha_1$ is emitted co-linearly with $\delta_1$, while if
$c_1=-1$ it is emitted contra-linearly. If $c_2=+1$ then $\alpha_2$ is
emitted contra-linearly with $\delta_2$, while if $c_2=-1$ it is
emitted co-linearly. The $\alpha_i$ particles in these extremal
configurations possess velocities given by
$p'(\alpha_i)/E'(\alpha_i)$. When $m(v_1)$ or $m(v_2)$ are maximised
or minimised these velocities define the coordinates at which the
boundaries of the annular regions in $\alpha_i$ phase space diagrams
such as Figure~\ref{fig1} cross the $\hat{x}$ axes. The invariant
masses of the visible particles in these extremal configurations can
be obtained by substituting the energies and momenta of $v_1$ and
$v_2$ from Eqns.~(\ref{eqn10}) into Eqn.~(\ref{eqn3}).

Now let us consider the kinematic configurations which generate
extremal values of $m(v_1,v_2)$. We shall label these configurations
with the signs of the $c_i$ quantities, which denote the directions of
the $\alpha_i$ particles in the $\delta_i$ rest frames, together with
the masses of the $v_i$ particles. The notation we shall use is of the
form $\{\pm_{v},\pm_{w}\}$, where the first(second) element inside the
brackets refers to particle 1(2), the signs correspond to the signs
of $c_i$, and the subscripts $v$ and $w$ describe the values of
respectively $m(v_1)$ and $m(v_2)$. So for example
$\{+_{\max},-_{\min}\}$ corresponds to a configuration in which
$\alpha_1$ is emitted co-linearly with $\delta_1$, $\alpha_2$ is
emitted co-linearly with $\delta_2$, $m(v_1)=\mmax(v_1)$ and
$m(v_2)=\mmin(v_2)$.

Using our new notation, the possible configurations maximising
$m(v_1,v_2)$ in the case of identical decay chains discussed in
Section~\ref{subsec2.1} are $\{-_{\min},+_{\min}\}$, corresponding to
Figure~\ref{fig1}(centre), and $\{-_{\max},+_{\max}\}$, corresponding
to Figure~\ref{fig1}(right). In the case of non-identical decay chains
we must also consider $\{-_{\max},+_{\min}\}$ and
$\{-_{\min},+_{\max}\}$. We have not exhausted the possibilities
however. In order to maximise $m(v_1,v_2)$ we must minimise both
$m(\alpha_1,\alpha_2)$ and the net momentum of the $\alpha_1\alpha_2$
system. In some models with highly asymmetric decay chains, for
example that represented in Figure~\ref{fig2}(right), these two
requirements are mutually exclusive -- decreasing
$m(\alpha_1,\alpha_2)$ increases the net invisible momentum and vice
versa. In such models we must find the values of $m(v_1)$ and $m(v_2)$
which maximise $m(v_1,v_2)$ when respectively $m(v_2)$ and $m(v_1)$
are maximised or minimised. These values are located at the turning
points of $m(v_1,v_2)$ and are given by:
\begin{multline}
\label{eqn11}
m^2_{\rm tp}(v_i) = m^2(\delta_i)+m^2(\alpha_i)\\-m(\alpha_i)\frac{2m^2(\delta_i)m(\delta_j)+AE(v_j)+Bp(v_j)}{\sqrt{m^2(\delta_j)[m^2(\delta_i)+m^2(v_j)]+m(\delta_j)[AE(v_j)+Bp(v_j)]}},
\end{multline}
where 
\begin{eqnarray}
\label{eqn11a}
A \equiv m^2(\omega)-m^2(\delta_1)-m^2(\delta_2), &\mbox{   }&
B \equiv \sqrt{A^2-4m^2(\delta_1)m^2(\delta_2)},
\end{eqnarray}
$E(v_j)$ and $p(v_j)$ are defined by Eqns.~(\ref{eqn7})
and~(\ref{eqn7a}), and $j\neq i$. Taking all possible combinations
into account this leads finally to eight possible values for
$\mmax(v_1,v_2)$ when Eqns.~(\ref{eqn7-1}) and~(\ref{eqn7-2}) are not
satisfied. These values are obtained from the configurations:
\begin{eqnarray}
\label{eqn12}
&\{-_{\min},+_{\min}\},\mbox{   }\{-_{\max},+_{\max}\},\mbox{   }\{-_{\min},+_{\max}\},\mbox{   }\{-_{\max},+_{\min}\},& \cr
&\{-_{\rm tp},+_{\min}\},\mbox{   }\{-_{\rm tp},+_{\max}\},\mbox{   }\{-_{\min},+_{\rm tp}\},\mbox{   }\{-_{\max},+_{\rm tp}\}.&
\end{eqnarray}
The value of $\mmax(v_1,v_2)$ is given by the maximum value of
Eqn.~(\ref{eqn3}) obtained by substituting the energies and momenta
from Eqns.~(\ref{eqn10}) for the eight configurations listed in
Eqn.~(\ref{eqn12}).

\section{Transverse mass end-points}\label{sec3}

The transverse mass $\mt$ of a set of visible and invisible decay
products can be calculated from their transverse momenta and
masses. The transverse momentum and mass of the aggregate visible
decay product $V$ of $\omega$ can be obtained by summing the
four-momenta of the aggregate visible decay products $v_1$ and $v_2$
of each decay chain, while the transverse momentum of the invisible
decay products is measured by the event $\etm$ vector. The optimum
definition of $\mt$ depends upon the value of the lower limit $\chi$
on the mass of the aggregate invisible decay product\footnote{Note
that this is not directly a limit on the mass of the individual
invisible decay products $\alpha_i$, for instance the LSPs in SUSY
models.}. If this limit is zero, for instance because the
$m(\alpha_i)$ are unknown, then the optimum definition is:
\begin{equation}
\label{eqn17}
\mt^2(0)\equiv m^2(V) + 2[E_T(V)\etm-p_x(V)p_x^{\rm
miss}-p_y(V)p_y^{\rm miss}],
\end{equation}
where 
\begin{equation}
E_T(V) \equiv \sqrt{p_T^2(V) + m^2(V)}.
\end{equation}
A configuration which maximises $\mt(0)$ is that in which $m(V)$ is
minimised and $p_T(V)$ is maximised. This requirement implies that
$\delta_1$, $\delta_2$, $\alpha_1$, $\alpha_2$, $v_1$ and $v_2$ must
all move in the laboratory transverse plane.

If the lower limit on the mass of the aggregate invisible decay
product is non-zero, for instance because the $m(\alpha_i)$ have been
measured, then the optimum definition is:
\begin{equation}
\label{eqn18}
\mt^2(\chi)\equiv m^2(V) + \chi^2 + 2[E_T(V)\etm(\chi)-p_x(V)p_x^{\rm
miss}-p_y(V)p_y^{\rm miss}],
\end{equation}
where
\begin{equation}
\etm(\chi) \equiv \sqrt{(\etm)^2 + \chi^2}.
\end{equation}
If the $m(\alpha_i)$ are known then a conservative value for $\chi$ is
$m(\alpha_1)+m(\alpha_2)$, which we shall use below. The absolute
maximum value of $\mt(m(\alpha_1)+m(\alpha_2))$ is $m(\omega)$, which
is obtained when the $\alpha_1$ is at rest with respect to $\alpha_2$,
and $\delta_1$, $\delta_2$, $\alpha_1$, $\alpha_2$, $v_1$ and $v_2$
are all moving in the laboratory transverse plane. Given the
discussion of Section~\ref{sec2}, we can use $\alpha_i$ phase space
diagrams such as Figures~\ref{fig1} and~\ref{fig2} to identify the
configurations which satisfy this requirement. Note however that
because we are now using only transverse momenta we must reinterpret
such diagrams as representing the transverse cartesian components of
velocity rather than cylindrical polar components about the $\omega$
decay axis. It is then clear that the configurations for which
$\mt(m(\alpha_1)+m(\alpha_2))=m(\omega)$ are those located in the
regions of the $\alpha_i$ phase space diagram in which the annular
regions overlap, provided all the motion lies in the
laboratory transverse plane. If the annular regions for a given model
do not overlap, or some of the particles move out of the laboratory
transverse plane, then this bound is not saturated.

\subsection{Identical chains}\label{subsec3.1}

In the case of identical decay chains $\mt(0)$ defined by
Eqn.~(\ref{eqn17}) can be maximised with the configurations
$\{+_{\min},+_{\min}\}$ and $\{-_{\min},-_{\min}\}$, which generate in
this case equal maxima. Substituting Eqns.~(\ref{eqn10}) into
Eqn.~(\ref{eqn17}) and assuming that all the motion lies in the
laboratory transverse plane we obtain the bound:
\begin{multline}
\label{eqn19}
\mtmax(0) = \frac{m(\omega)}{2m^2(\delta)}\Big[m^2(\delta)-m^2(\alpha)+\mmin^2(v)+\\\sqrt{\big[m^2(\delta)-m^2(\alpha)+\mmin^2(v)\big]^2-4m^2(\delta)\mmin^2(v)}\Big],
\end{multline}
where $\mmin(v)$ is the minimum value of the invariant mass of the
{\it individual} aggregate visible decay products $v_1$ and $v_2$, as
in Section~\ref{sec2}. If $\mmin(v)=0$ this reduces to
\begin{equation}
\label{eqn20}
\mtmax(0) = m(\omega)\Big[1-\frac{m^2(\alpha)}{m^2(\delta)}\Big],
\end{equation}
which is always less than $m(\omega)$. 

In the case of $\mt(m(\alpha_1)+m(\alpha_2))$, i.e. $\mt(2m(\alpha))$,
the condition that $\mtmax(2m(\alpha))=m(\omega)$ is equivalent to
requiring
\begin{equation}
\label{eqn21}
m^2(\delta)-m(\alpha)[m(\omega)-m(\alpha)] \geq \mmin^2(v),
\end{equation}
which is a less stringent requirement than Eqn.~(\ref{eqn5}). Models
satisfying this requirement include those represented in both
Figure~\ref{fig1}(left) and Figure~\ref{fig1}(right). In the latter
case the configurations saturating the bound possess $\alpha_i$
particles moving with $\beta_x=0$ and $\beta_y \neq 0$, in other words
emitted transverse to the $\delta_i$ direction in the $\omega$ rest
frame. The fact that for a given model multiple configurations can
saturate the bound rather than just one (as is the case for
$m(v_1,v_2)$) shows that in principle the $\mt(2m(\alpha))$ end-point
can be more prominent, as shall be discussed in Section~\ref{sec4}.

If the requirement Eqn.~(\ref{eqn21}) is not satisfied, as is the case
for the model represented in Figure~\ref{fig1}(centre), then the
configuration which maximises $\mt(2m(\alpha))$ is that which
minimises $m(\alpha_1,\alpha_2)$, with all the motion in the
laboratory transverse plane. This configuration is
$\{-_{\min},+_{\min}\}$, which is also that configuration which
maximises $m(v_1,v_2)$ given $m(v)=\mmin(v)$ (see
Section~\ref{subsec2.1}). In this case the aggregate invisible and
visible transverse momenta are zero and thus from Eqn.~(\ref{eqn18}):
\begin{equation}
\label{eqn22-1}
\mtmax(2m(\alpha)) = 2m(\alpha) + m(v_1,v_2),
\end{equation}
or equivalently from Eqn.~(\ref{eqn6}):
\begin{multline}
\label{eqn22}
\mtmax(2m(\alpha)) = 2m(\alpha) + \frac{m(\omega)}{2m^2(\delta)}\Big[m^2(\delta)-m^2(\alpha)+\mmin^2(v)+\\\frac{1}{m(\omega)}\sqrt{\big(m^2(\omega)-4m^2(\delta)\big)\big(\big[m^2(\delta)-m^2(\alpha)+\mmin^2(v)\big]^2-4m^2(\delta)\mmin^2(v)\big)}\Big].
\end{multline}

\subsection{Non-identical chains}\label{subsec3.2}

In the case of non-identical decay chains the analysis for $\mt(0)$ is
very similar to that in the case of identical decay
chains. Configurations which can generate the maximum values of
$\mt(0)$ are again $\{+_{\min},+_{\min}\}$ and
$\{-_{\min},-_{\min}\}$, although in this case they may in principle
generate different values of $\mtmax(0)$. Assuming that all the motion
lies in the laboratory transverse plane, these configurations can be
used with Eqns.~(\ref{eqn10}) and~(\ref{eqn17}) to generate two
possible values of $\mtmax(0)$ with the maximum of these two values
used.

When considering $\mt(m(\alpha_1)+m(\alpha_2))$ the analysis is more
complicated. To simplify the discussion we shall first define decay
chain 1 to be that chain which generates the smaller annular region in
the $\alpha_i$ phase space diagram (see e.g. Figure~\ref{fig2}). In
other words we label decay chains such that the difference in $\beta$
values for $\alpha_1$ configurations $+_{\min}$ and $-_{\min}$ is less
than the difference in $\beta$ values for $\alpha_2$ configurations
$+_{\min}$ and $-_{\min}$. Given that we have already defined the
$+\hat{x}$ direction to be that direction in which $\delta_1$ is
emitted in the $\omega$ rest frame, this additional definition then
maps Figure~\ref{fig2}(centre) onto Figure~\ref{fig2}(left).

With this new definition of decay chains 1 and 2 we find that the
requirement that the two annular regions in the $\alpha_i$ phase space
diagram overlap is equivalent to requiring that the $\beta$ value for
the $\alpha_1$ configuration $-_{\min}$ is less than the $\beta$ value
for the $\alpha_2$ configuration $+_{\min}$, and the $\beta$ value for
the $\alpha_1$ configuration $+_{\min}$ is greater than the $\beta$
value for the $\alpha_2$ configuration $+_{\max}$. These requirements
are not satisfied by any of the three models represented in
Figure~\ref{fig2} and in such cases the maximum value of
$\mt(m(\alpha_1)+m(\alpha_2))$ is obtained from one of three
configurations: $\{-_{\min},+_{\min}\}$ (as in the case of identical
chains), $\{-_{\max},-_{\min}\}$ and $\{+_{\min},+_{\max}\}$. The
latter two configurations map to each other if the decay chains are
defined as described above.

\section{Example: SUSY higgs decaying to gauginos at the LHC}\label{sec4}

\FIGURE[t]{
\epsfig{file=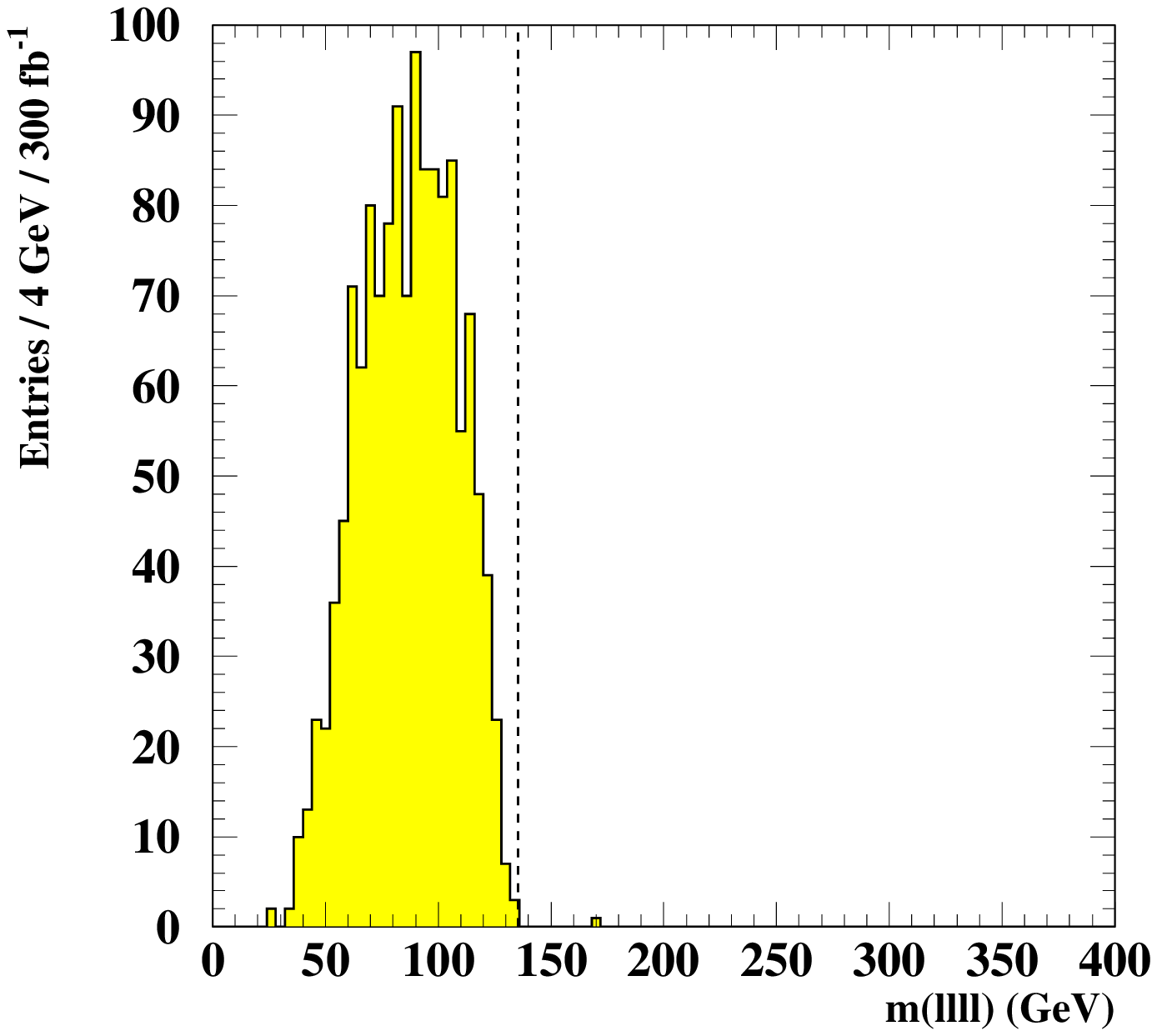,height=3.5in}
\caption{\label{fig3} Distribution of detector-level four-lepton
invariant mass values for $H/A\ra \chioii\chioii \ra \ell\ell\ell\ell
\chioi\chioi$ events from the {\it Point A} model. The dashed vertical
line represents the expected end-point position from
Eqn.~(\ref{eqn4}). The small population of events lying beyond the
expected end-point is generated by detector mis-measurement. }}

\FIGURE[t]{
\epsfig{file=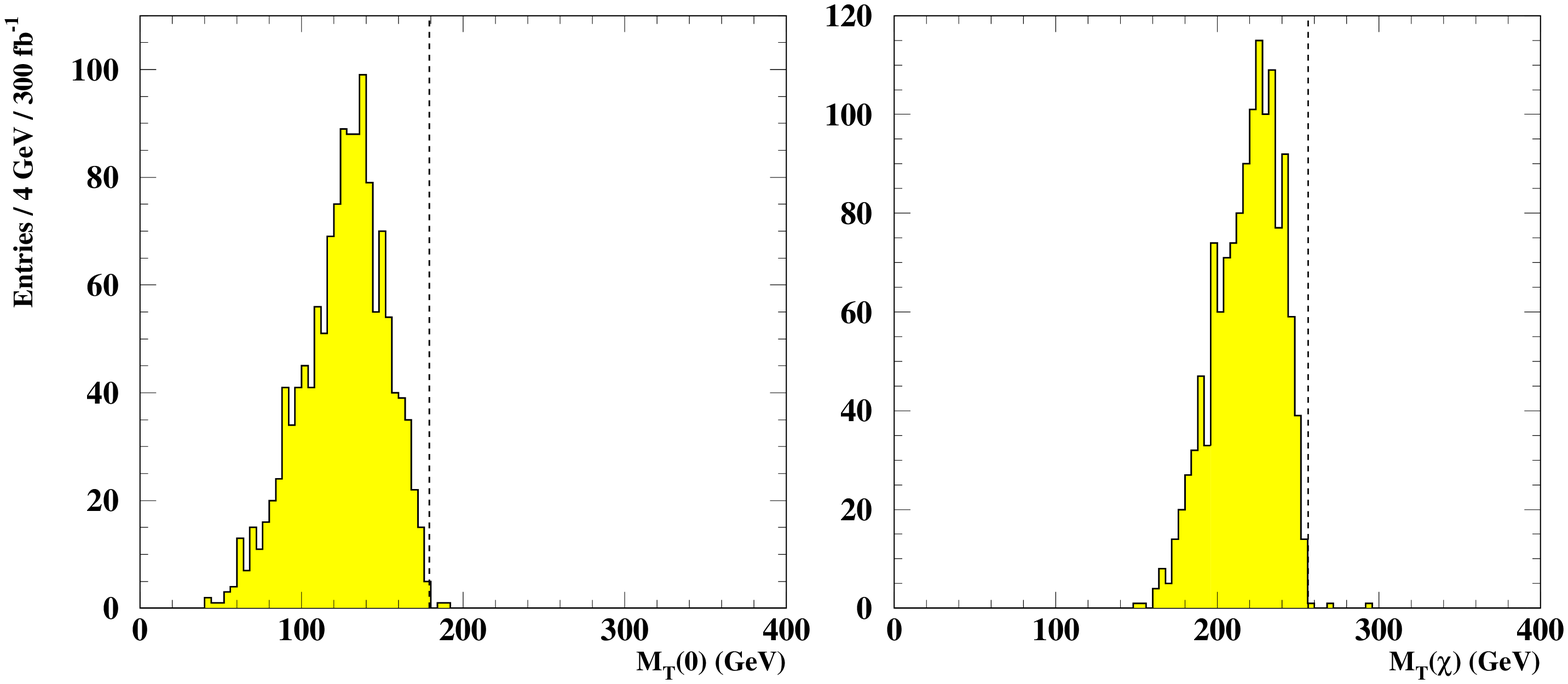,height=3.0in}
\caption{\label{fig4} Distributions of detector-level transverse mass
values for $H/A\ra \chioii\chioii \ra \ell\ell\ell\ell \chioi\chioi$
events the {\it Point A} model. In the left(right)-hand figure the
transverse mass is defined by Eqn.~(\ref{eqn17})
(Eqn.~(\ref{eqn18})). The dashed vertical lines represent the expected
end-point positions, given by respectively Eqn.~(\ref{eqn20}) and
$m(\omega)$. The small populations of events lying beyond the expected
end-point are generated by detector mis-measurement.}}

In order to demonstrate some of the simpler aspects of the above
discussion a Monte Carlo study was conducted of heavy SUSY higgs
particles decaying to the four lepton plus $\etm$ final state via
$\chioii\chioii$ with three-body $\chioii$ lepton producing decays:
\begin{equation}
\label{eqn23}
H/A \ra \chioii\chioii \ra \ell\ell\ell\ell \chioi\chioi.
\end{equation}
The model chosen for study was the mSUGRA {\it Point A} model from
Ref.~\cite{Bisset:2007mi}, for which $m(H/A)=256$ GeV,
$m(\chioii)=110$ GeV and $m(\chioii)=60$ GeV. The {\tt ISASUGRA 7.69}
\cite{Paige:2003mg} RGE code coupled to {\tt HDECAY}
\cite{Djouadi:1997yw} was used to generate the sparticle mass spectra
and branching ratios, while {\tt HERWIG 6.510}
\cite{Corcella:2000bw,Moretti:2002eu} and {\tt ACERDET}
\cite{RichterWas:2002ch} were used to generate and simulate 14 TeV LHC
events corresponding to about 300 fb$^{-1}$ of data (assuming perfect
identification efficiency). Detailed studies of SM and SUSY
backgrounds to the $4\ell+\etm$ channel analysis are beyond the scope
of this paper -- for this simple illustrative study only signal events
were considered with a rudimentary detector level event selection
requiring merely the presence of four isolated leptons in
opposite-sign same-flavour pairs. The resulting values of
$m(v_1,v_2)=m(\ell_1,\ell_2,\ell_4,\ell_4)$, $\mt(0)$ and
$\mt(2m(\chioi))$ are shown in Figures~\ref{fig3} and~\ref{fig4}.

In the case of $H/A\ra \chioii\chioii \ra \ell\ell\ell\ell
\chioi\chioi$ events the two $\delta_i$ decay chains are identical, with $\omega
\equiv H/A$, $\delta_1=\delta_2=\chioii$, $\alpha_1=\alpha_2=\chioi$
and $v_1=v_2=\ell^+\ell^-$. In the example considered here
$\mmax(v_i)=m(\chioii)-m(\chioi)$ due to the three-body nature of the
$\chioii$ decays, while $\mmin(v_i)=0$. Consequently the requirement
in Eqn.~(\ref{eqn5}) is satisfied and hence
$\mmax(v_1,v_2)=m(H/A)-2m(\chioi)=136$ GeV. The requirement in
Eqn.~(\ref{eqn21}) is also satisfied giving $\mtmax(2m(\chioi))=256$
GeV. The $\mt(0)$ end-point is given by Eqn.~(\ref{eqn20}) leading to
$\mtmax(0)=180$ GeV. These expected end-point positions are
represented in Figures~\ref{fig3} and~\ref{fig4} by vertical dashed
lines and agree well with the observed end-points. It is interesting
to note that the larger number of configurations saturating the bound
on $\mt(2m(\chioi))$ compared with $m(v_1,v_2)$ in this case leads to
a more prominent end-point with a steeper gradient. This is true even
at detector-level following smearing of the event $\etm$ values used
to calculate $\mt(2m(\chioi))$.

\section{Conclusions}\label{sec5}

This brief paper has discussed the positions of end-points in the
invariant mass and transverse mass distributions of the decay products
of heavy particles decaying to pairs of semi-invisibly decaying
products. The formulae presented here may prove useful for mass
measurements if SUSY higgs bosons decaying to gauginos are observed at
the LHC. The same techniques may also prove useful in other new
physics scenarios, for instance given heavy states decaying via pairs
of new $W'$ bosons to massive stable right-handed neutrinos.

\section*{Acknowledgements}
DRT wishes to thank Alan Barr, Ben Gripaios, Chris Lester and Giacomo
Polesello for helpful comments. DRT wishes to acknowledge STFC and the
Leverhulme Trust for support.

\end{document}